\documentstyle[twocolumn,aps,epsf]{revtex}    

\begin{document}
\draft
\noindent
{\bf Comment on \lq Missing 2$k_F$ Response for Composite Fermions in Phonon
Drag\rq }\\

In a recent Letter Zelakiewicz {\sl et al.}~report on the mysterious
absence of a Kohn anomaly in the phonon mediated drag of composite
fermions (CFs) in double layer electron systems [1].
In particular the temperature dependence and the magnitude of the drag
transresistivity when moving away from $\nu = 1/2$ are claimed
to be inconsistent with current expectations for CFs.
Here we show that all the results in [1] can be simply explained
by phonon drag of free electrons in the lowest Landau level [2].
In this model the electron-phonon interaction follows a universal behavior
when plotted as a function of $l_B/\lambda_p$ where
$\lambda_p =  \hbar v / k_B T$ is the typical phonon wave length for
acoustic phonons with a sound velocity $v$
and $l_B = (\hbar/eB)^{1/2}$ is the magnetic length.
This universality entails a scaling of the phonon mediated drag
$\rho_D (T,B) = \rho_s (\nu) f (l_B/\lambda_p)$ with
a filling factor dependent normalization factor $\rho_s (\nu)$ 
and a scaling function $f(l_B/\lambda_p) $.

In Fig.~1 the results of such an analysis are shown.
The drag transresistivity $\rho_D$ as presented in [1], arbitrarily
normalized to its value $\rho_s$ at $T B^{-1/2} = 0.6$~TK$^{-1/2}$,
is plotted as a function of $T B^{-1/2} (\propto l_B/\lambda_p)$.
Indeed,  as indicated by the lines,
all the drag data presented in [1] scale to a single function
in particular around the position where a maximum occurs in $\rho_D/T^2$.
The identical dependence of
$\rho_s$ on $B$ and the inverse electron concentration $1/n$
(insets in Fig.~1) show that $\rho_s$ indeed only depends
on $\nu=hn/eB$ supplying an additional indication for
the validity of a free electron model for all the data
presented in [1].

It was indicated in [1] that such a simple rescaling does not seem
to work for data at $\nu=1/4$, where a 10\% {\sl lower} position of the
\lq Kohn-anomaly maximum\rq~was observed. We note, however, that these
data are taken at a considerably lower electron concentration. On the
other hand, when extrapolating the data from Fig.~2a of [1] to lower
fields, the maximum for $\nu=3/4$ definitely occurs at much lower $T$
than that for $\nu=1/2$. Therefore it seems that the data obtained on the
$1/4$-$3/4$ CF-family are far from being conclusive to discriminate
between a CF model and a free electron picture.

Moreover, it is worthwhile  mentioning that
experiments on the {\sl direct} phonon drag of CFs measured by
thermopower (TEP) showed that
any CF signature disappears as soon as the
related minima in $\rho_{xx}$ at odd denominator filling factors start
to weaken [3].
At the high temperatures as used in [1]
the phonon drag TEP is well described by the model of
non-interacting Landau-quantized electrons [4].
Of course it can not be excluded that CFs still exist
at these high temperatures. However, no specific property
which can not be simply related to free electrons is
observed in phonon drag and a CF interpretation becomes meaningless.
The fact that even the $\nu=1/3$ and $\nu=2/3$ minima
(inset of Fig.~2a in [1])
are merely visible in $\rho_D$ and very weakly developed in $\rho_{xx}$
strongly proposes that it is sufficient to use a model as
in [4] for a proper analysis of the data in [1].

In conclusion we have shown that all the data on the phonon mediated drag
in coupled two-dimensional electron systems in high magnetic fields
presented in [1] can be straightforwardly
explained in a framework of electrons in the lowest Landau level
without the need to use any new CF models.\\

\noindent
U.~Zeitler$^1$ and J.~G.~S.~Lok$^2$\\
\vspace{-1.6em} 
 \begin{small}
\begin{tabbing}
$^1$ \=Institut f\"ur Festk\"orperphysik, Universit\"at Hannover,\\
     \>Appelstra{\ss}e 2, 30167 Hannover, Germany.\\
$^2$ \>Max-Planck-Institut f\"ur Festk\"orperforschung,\\
     \>Heisenbergstr.~1, 70569 Stuttgart, Germany.

\end{tabbing}
\end{small}

\noindent
Received: February 23, 2001 \newline

\noindent
[1] S.~Zelakiewicz {\sl et al.},
   Phys.~Rev.~Lett.~{\bf 85}, 1942 (2000).\newline
[2] V.~I.~Fal'ko and S.~V.~Iordanskii,\newline
\hspace*{1.2em} J.~Phys.: Condens.~Matter {\bf 4}, 9201 (1992).\newline
[3] B.~Tieke {\sl et al.}, Phys.~Rev.~B {\bf 58}, 2017 (1998),\newline
\hspace*{1.2em}      Phys.~Rev.~Lett.~{\bf 76}, 3630 (1996).\newline
[4] U.~Zeitler {\sl et al.}, Surf.~Sci.~{\bf 305}, 91 (1994). \newline

\begin{figure}[b]
\vspace*{1em}   
\centerline{\epsfxsize=7cm
\epsfbox{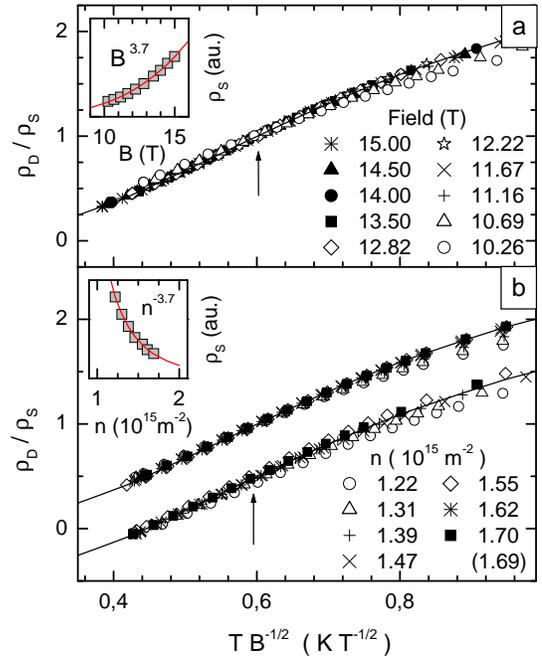}}
\vspace*{1em} 
\caption{Dependence of the normalized drag transresistivity $\rho_D / \rho_s$
on $T B^{-1/2}$ when varying the magnetic field at constant density
$n = 1.55 \times 10^{15}$~m$^{-2}$ (a), varying the density at
constant $B=12.82$~T ((b), top trace) and varying both $B$ and $n$
at constant Landau level filling $\nu=1/2$ ((b), bottom trace,
shifted for clarity by -0.5).
The data are extracted from Fig.~2a, Fig.~2b and Fig.~3 of [1] using
the same symbols.
The arrows mark the position where $\rho_D/T^2$ reaches its maximum.
The insets show the dependence of $\rho_s$ on
$B$ at constant $n$ (a) and on $n$ at constant $B$ (b).
}
\end{figure}

\end{document}